\documentclass[fleqn,11pt]{article}

\usepackage{latexsym,ifthen,epsfig,color}
\usepackage{amsmath,amssymb,amsthm}
\usepackage{float}

\newcommand{\join}{\text{\textcircled{{\footnotesize 1}}}}
\newcommand{\cojoin}{\text{\textcircled{{\footnotesize 0}}}}

\newcommand{\NP}{\ensuremath{\mathbb{NP}}}

\newtheorem{theorem}{Theorem}
\newtheorem{lemma}{Lemma}

\newtheorem{corollary}{Corollary}

\newtheorem{observation}{Observation}

\newtheorem{proc}{Procedure}[section]

\begin{document}

\title{Finding Dominating Induced Matchings in $S_{1,1,5}$-Free Graphs in Polynomial Time}

\author{
Andreas Brandst\"adt\footnote{Institut f\"ur Informatik,
Universit\"at Rostock, A.-Einstein-Str.\ 22, D-18051 Rostock, Germany,
{\texttt andreas.brandstaedt@uni-rostock.de}}
\and
Raffaele Mosca\footnote{Dipartimento di Economia, Universit\'a degli Studi ``G.\ D'Annunzio''
Pescara 65121, Italy.
{\texttt r.mosca@unich.it}}
}

\maketitle

\begin{abstract}
Let $G=(V,E)$ be a finite undirected graph. An edge set $E' \subseteq E$ is a {\em dominating induced matching} ({\em d.i.m.}) in $G$ if every edge in $E$ is intersected by exactly one edge of $E'$.
The \emph{Dominating Induced Matching} (\emph{DIM}) problem asks for the existence of a d.i.m.\ in $G$; this problem is also known as the \emph{Efficient Edge Domination} problem; it is the Efficient Domination problem for line graphs.

The DIM problem is \NP-complete even for very restricted graph classes such as planar bipartite graphs with maximum degree 3 but is solvable in linear time for $P_7$-free graphs, and in polynomial time for $S_{1,2,4}$-free graphs as well as for $S_{2,2,2}$-free graphs and for $S_{2,2,3}$-free graphs. In this paper, combining two distinct approaches, we solve it in polynomial time for $S_{1,1,5}$-free graphs.
\end{abstract}


\section{Introduction}\label{sec:intro}

Let $G=(V,E)$ be a finite undirected graph. A vertex $v \in V$ {\em dominates} itself and its neighbors. A vertex subset $D \subseteq V$ is an {\em efficient dominating set} ({\em e.d.s.} for short) of $G$ if every vertex of $G$ is dominated by exactly one vertex in $D$.
The notion of efficient domination was introduced by Biggs \cite{Biggs1973} under the name {\em perfect code}.
The {\sc Efficient Domination} (ED) problem asks for the existence of an e.d.s.\ in a given graph $G$ (note that not every graph has an e.d.s.)

A set $M$ of edges in a graph $G$ is an \emph{efficient edge dominating set} (\emph{e.e.d.s.} for short) of $G$ if and only if it is an e.d.s.\ in its line graph $L(G)$. The {\sc Efficient Edge Domination} (EED) problem asks for the existence of an e.e.d.s.\ in a given graph $G$. Thus, the EED problem for a graph $G$ corresponds to the ED problem for its line graph $L(G)$. Note that not every graph has an e.e.d.s. An efficient edge dominating set is also called \emph{dominating induced matching} ({\em d.i.m.} for short), and the EED problem is called the {\sc Dominating Induced Matching} (DIM) problem in various papers (see e.g. \cite{BraHunNev2010,BraMos2014,BraMos2017,BraMos2017/2,BraMos2017/3,CarKorLoz2011,HerLozRieZamdeW2015,KorLozPur2014}); subsequently, we will use this notation instead of EED.

In \cite{GriSlaSheHol1993}, it was shown that the DIM problem is \NP-complete; see also~\cite{BraHunNev2010,CarKorLoz2011,LuKoTan2002,LuTan1998}.
However, for various graph classes, DIM is solvable in polynomial time. For mentioning some examples, we need the following notions:

Let $P_k$ denote the chordless path $P$ with $k$ vertices, say $a_1,\ldots,a_k$, and $k-1$ edges $a_ia_{i+1}$, $1 \le i \le k-1$; we also denote it as $P=(a_1,\ldots,a_k)$.

For indices $i,j,k \ge 0$, let $S_{i,j,k}$ denote the graph $H$ with vertices $u,x_1,\ldots,x_i$, $y_1,\ldots,y_j$, $z_1,\ldots,z_k$ such that the subgraph induced by $u,x_1,\ldots,x_i$ forms a $P_{i+1}$ $(u,x_1,\ldots,x_i)$, the subgraph induced by $u,y_1,\ldots,y_j$ forms a $P_{j+1}$ $(u,y_1,\ldots,y_j)$, and the subgraph induced by $u,z_1,\ldots,z_k$ forms a $P_{k+1}$ $(u,z_1,\ldots,z_k)$, and there are no other edges in $S_{i,j,k}$; $u$ is called the {\em center} of $H$.
Thus, {\em claw} is $S_{1,1,1}$, and $P_k$ is isomorphic to $S_{k-1,0,0}$.

For a set ${\cal F}$ of graphs, a graph $G$ is called {\em ${\cal F}$-free} if no induced subgraph of $G$ is contained in ${\cal F}$.
If $|{\cal F}|=1$, say ${\cal F}=\{H\}$, then instead of $\{H\}$-free, $G$ is called $H$-free.

\medskip

The following results are known:

\begin{theorem}\label{DIMpolresults}
DIM is solvable in polynomial time for
\begin{itemize}
\item[$(i)$]  $S_{1,1,1}$-free graphs $\cite{CarKorLoz2011}$,
\item[$(ii)$] $S_{1,2,3}$-free graphs $\cite{KorLozPur2014}$,
\item[$(iii)$] $S_{2,2,2}$-free graphs $\cite{HerLozRieZamdeW2015}$,
\item[$(iv)$] $S_{1,2,4}$-free graphs $\cite{BraMos2017/2}$,
\item[$(v)$] $S_{2,2,3}$-free graphs $\cite{BraMos2017/3}$,
\item[$(vi)$]  $P_7$-free graphs $\cite{BraMos2014}$ (in this case even in linear time),
\item[$(vii)$] $P_8$-free graphs $\cite{BraMos2017}$.
\end{itemize}
\end{theorem}

In \cite{HerLozRieZamdeW2015}, it is conjectured that for every fixed $i,j,k$, DIM is solvable in polynomial time for $S_{i,j,k}$-free graphs (actually, an even stronger conjecture is mentioned in \cite{HerLozRieZamdeW2015}); this includes $P_k$-free graphs for $k \ge 9$.

\medskip

In this paper we show that DIM can be solved in polynomial time for $S_{1,1,5}$-free graphs (generalizing the corresponding result for $P_7$-free graphs). The approach is based on that described in \cite{BraMos2017}, i.e. by fixing an edge in the possible e.d.s. and studying the corresponding distance levels, and is developed thanks to that described in \cite{HerLozRieZamdeW2015,KorLozPur2014}, i.e., by seeing the problem in terms of ``coloring'' and defining polynomially many families of feasible partial colorings. The proposed solution considers a sequence of more and more general instances, with respect to the assumption that distance levels could be empty, until to consider the general case. In particular it seems that in view of a possible extension of the result for $S_{1,1,k}$-free graphs, for $k \geq 6$, the main obstacle would be that of solving the very first instances of the sequence.

\section{Definitions and Basic Properties}\label{sec:basicnotionsresults}

\subsection{Basic notions}\label{subsec:basicnotions}

Let $G$ be a finite undirected graph without loops and multiple edges. Let $V(G)$ or $V$ denote its vertex set and $E(G)$ or $E$ its edge set; let $n=|V|$ and $m=|E|$.
For $v \in V$, let $N(v):=\{u \in V: uv \in E\}$ denote the {\em open neighborhood of $v$}, and let $N[v]:=N(v) \cup \{v\}$ denote the {\em closed neighborhood of $v$}. If $xy \in E$, we also say that $x$ and $y$ {\em see each other}, and if $xy \not\in E$, we say that $x$ and $y$ {\em miss each other}.
A vertex set $S$ is {\em independent} in $G$ if for every pair of vertices $x,y \in S$, $xy \not\in E$. A vertex set $Q$ is a {\em clique} in $G$ if for every pair of vertices $x,y \in Q$, $x \neq y$, $xy \in E$. For $uv \in E$ let $N(uv):= N(u) \cup N(v) \setminus \{u,v\}$ and $N[uv]:= N[u] \cup N[v]$.

For $U \subseteq V$, let $G[U]$ denote the subgraph of $G$ induced by vertex set $U$. Clearly $xy \in E$ is an edge in $G[U]$ exactly when $x \in U$ and $y \in U$; thus, $G[U]$ can simply be denoted by $U$ (if understandable).

For $A \subseteq V$ and $B \subseteq V$, $A \cap B = \emptyset$, we say that: $A \cojoin B$ if each vertex of $A$ misses each vertex of $B$; $A \join B$ if each vertex of $A$ sees each vertex of $B$; $A$ {\em contacts} $B$ if some vertex of $A$ sees some vertex of $B$.
For $A=\{a\}$, we simply denote $A \join B$ by $a \join B$, and correspondingly $A \cojoin B$ by $a \cojoin B$, and correspondingly say that $a$ {\em contacts} $B$. If for $A' \subseteq A$, $A' \cojoin (A \setminus A')$, we say that $A'$ is {\em isolated} in $G[A]$. 

For graphs $H_1$, $H_2$ with disjoint vertex sets, $H_1+H_2$ denotes the disjoint union of $H_1$, $H_2$, and for $k \ge 2$, $kH$ denotes the disjoint union of $k$ copies of $H$. For example, $2P_2$ is the disjoint union of two edges.

As already mentioned, a {\em chordless path} $P_k$, $k \ge 2$, has $k$ vertices, say $v_1,\ldots,v_k$, and $k-1$ edges $v_iv_{i+1}$, $1 \le i \le k-1$;
the {\em length of $P_k$} is $k-1$. We also denote it as $P=(v_1,\ldots,v_k)$.

A {\em chordless cycle} $C_k$, $k \ge 3$, has $k$ vertices, say $v_1,\ldots,v_k$, and $k$ edges $v_iv_{i+1}$, $1 \le i \le k-1$, and $v_kv_1$; the {\em length of $C_k$} is $k$.

Let $K_i$, $i \ge 1$, denote the clique with $i$ vertices. Let $K_4-e$ or {\em diamond} be the graph with four vertices, say $v_1,v_2,v_3,u$, such that $(v_1,v_2,v_3)$ forms a $P_3$ and $u \join \{v_1,v_2,v_3\}$; its {\em mid-edge} is the edge $uv_2$.

A {\em butterfly} has five vertices, say, $v_1,v_2,v_3,v_4,u$, such that $v_1,v_2,v_3,v_4$ induce a $2P_2$ with edges $v_1v_2$ and $v_3v_4$ (the {\em peripheral edges} of the butterfly), and $u \join \{v_1,v_2,v_3,v_4\}$.

We often consider an edge $e = uv$ to be a set of two vertices; then it makes sense to say, for example, $u \in e$ and $e \cap e' \neq \emptyset$, for an edge $e'$. For two vertices $x,y \in V$, let $dist_G(x,y)$ denote the {\em distance between $x$ and $y$ in $G$}, i.e., the length of a shortest path between $x$ and $y$ in $G$.
The {\em distance between a vertex $z$ and an edge $xy$} is the length of a shortest path between $z$ and $x,y$, i.e., $dist_G(z,xy)= \min\{dist_G(z,v): v \in \{x,y\}\}$.
The {\em distance between two edges} $e,e' \in E$ is the length of a shortest path between $e$ and $e'$, i.e., $dist_G(e,e')= \min\{dist_G(u,v): u \in e, v \in e'\}$.
In particular, this means that $dist_G(e,e')=0$ if and only if $e \cap e' \neq \emptyset$.

An edge subset $M \subseteq E$ is an {\em induced matching} if the pairwise distance between its members is at least 2, that is, $M$ is isomorphic to $kP_2$ for $k=|M|$. Obviously, if $M$ is a d.i.m.\ then $M$ is an induced matching.

Clearly, $G$ has a d.i.m.\ if and only if every connected component of $G$ has a d.i.m.; from now on, connected components are mentioned as {\em components}.

Note that if $G$ has a d.i.m.\ $M$, and $V(M)$ denotes the vertex set of $M$ then $V \setminus V(M)$ is an independent set, say $I$, i.e.,
\begin{equation}\label{IV(M)partition}
V \mbox{ has the partition } V = V(M) \cup I .
\end{equation}

From now on, all vertices in $I$ are colored white and all vertices in $V(M)$ are colored black. According to \cite{HerLozRieZamdeW2015}, we also use the following notions: A partial black-white coloring of $V(G)$ is {\em feasible} if the set of white vertices is an independent set in $G$ and every black vertex has at most one black neighbor. A complete black-white coloring of $V(G)$ is {\em feasible} if the set of white vertices is an independent set in $G$ and every black vertex has exactly one black neighbor. Clearly, $M$ is a d.i.m.\ of $G$ if and only if the black vertices $V(M)$ and the white vertices $V \setminus V(M)$ form a complete feasible coloring of $V(G)$.

\subsection{Reduction steps, forbidden subgraphs, forced edges, and excluded edges}\label{forbidsubgrforcededges}

Various papers on this topic introduced and applied some {\em forcing rules} for reducing the graph $G$ to a subgraph $G'$ such that $G$ has a d.i.m.\ if and only if $G'$ has a d.i.m., based on the condition that for a d.i.m.\ $M$, $V$ has the partition $V= V(M) \cup I$ such that all vertices in $V(M)$ are black and all vertices in $I$ are white (recall (\ref{IV(M)partition})).

\medskip

A vertex $v \in V$ is {\em forced to be black} if for every d.i.m.\ $M$ of $G$, $v \in V(M)$.
Analogously, a vertex $v \in V$ is {\em forced to be white} if for every d.i.m.\ $M$ of $G$, $v \notin V(M)$.

Clearly, if $uv \in E$ and if $u,v$ are forced to be black, then $uv$ is contained in every (possible) d.i.m. of $G$.

\medskip

An edge $e \in E$ is a {\em forced} edge of $G$ if for every d.i.m.\ $M$ of $G$,
$e \in M$. Analogously, an edge $e \in E$ is an {\em excluded} edge of $G$ if for every d.i.m.\ $M$ of $G$,
$e \not \in M$.

\medskip

For the correctness of the reduction steps, we have to argue that $G$ has a d.i.m.\ if and only if the reduced graph $G'$ has one (provided that no contradiction arises in the vertex coloring, i.e., it is feasible).

Then let us introduce two reduction steps which will be applied later.

\medskip

\noindent
{\bf Vertex Reduction.} Let $u \in V(G)$. If $u$ is forced to be white, then
\begin{itemize}
\item[$(i)$] color black all neighbors of $u$, and
\item[$(ii)$] remove $u$ from $G$.
\end{itemize}

Let $G'$ be the reduced subgraph. Clearly, Vertex Reduction is correct, i.e., $G$ has a d.i.m.\ if and only if $G'$ has a d.i.m. 

\medskip

\noindent
{\bf Edge Reduction.} Let $uv \in E(G)$. If $u$ and $v$ are forced to be black, then
\begin{itemize}
\item[$(i)$] color white all neighbors of $u$ and of $v$ (other than $u$ and $v$), and
\item[$(ii)$] remove $u$ and $v$ from $G$.
\end{itemize}

Again, clearly, Edge Reduction is correct, i.e., $G$ has a d.i.m.\ if and only if the reduced subgraph $G'$ has a d.i.m. \\

The subsequent notions and observations lead to some possible reductions (some of them are mentioned e.g.\ in \cite{BraHunNev2010,BraMos2014,BraMos2017}).

\begin{observation}[\cite{BraHunNev2010,BraMos2014,BraMos2017}]\label{dimC3C5C7C4}
Let $M$ be a d.i.m.\ of $G$.
\begin{itemize}
\item[$(i)$] $M$ contains at least one edge of every odd cycle $C_{2k+1}$ in $G$, $k \ge 1$,
and exactly one edge of every odd cycle $C_3$, $C_5$, $C_7$ in $G$.
\item[$(ii)$] No edge of any $C_4$ can be in $M$.
\item[$(iii)$] For each $C_6$ either exactly two or none of its edges are in $M$.
\end{itemize}
\end{observation}

\noindent
{\bf Proof.} See e.g.\ Observation 2 in \cite{BraMos2014}.

\medskip

In what follows, we will also refer to Observation \ref{dimC3C5C7C4}~$(i)$ (with respect to $C_3$) as to the {\em triangle-property}, and to Observation \ref{dimC3C5C7C4}~$(ii)$ as to the {\em $C_4$-property}.

\medskip

Since by Observation \ref{dimC3C5C7C4} $(i)$, every triangle contains exactly one $M$-edge, and the pairwise distance of $M$-edges is at least 2, we have:

\begin{corollary}\label{cly:K4free}
If $G$ has a d.i.m.\ then $G$ is $K_4$-free.
\end{corollary}

{\sc Assumption 1.} From now on, by Corollary \ref{cly:K4free}, we assume that the input graph is $K_4$-free (else it has no d.i.m.).

\medskip

Clearly, it can be checked (directly) in polynomial time whether the input graph is $K_4$-free.

\medskip

Recall that a d.i.m.\ $M$ in $G$ is an induced matching, and the distance between every pair of edges in $M$ is at least 2 (which is the {\em distance property}).
By Observation \ref{dimC3C5C7C4} $(i)$ with respect to $C_3$ and the distance property, we have the following:

\begin{observation}\label{obs:diamondbutterfly}
The mid-edge of any induced diamond in $G$ and the two peripheral edges of any induced butterfly in $G$ are forced edges of $G$.
\end{observation}

{\sc Assumption 2.} From now on, by Observation \ref{obs:diamondbutterfly}, we assume that the input graph is (diamond,butterfly)-free.

\medskip

In particular, we can apply the Edge Reduction to each mid-edge of any induced diamond and to each peripheral edge of any induced butterfly; that can be done in polynomial time.

\subsection{The distance levels of an $M$-edge $xy$ in a $P_3$}\label{subsec:distlevels}

If for $xy \in E$, an edge $e \in E$ is contained in {\bf every} d.i.m.\ $M$ of $G$ with $xy \in M$, we say that $e$ is an {\em $xy$-forced} $M$-edge, and analogously, if an edge $e \in E$ is contained in {\bf no} d.i.m.\ $M$ of $G$ with $xy \in M$, we say that $e$ is {\em $xy$-excluded}. The Edge Reduction for forced edges can also be applied for $xy$-forced edges (then, in the unsuccessful case, $G$ has no d.i.m.\ containing $xy$), and correspondingly for $xy$-forced white vertices (resulting from the black color of $x$ and $y$), the Vertex Reduction can be applied.

\medskip

Based on \cite{BraMos2017}, we first describe some general structure properties for the distance levels of an edge in a d.i.m.\ $M$ of $G$.
Since $G$ is $(K_4$, diamond, butterfly)-free, we have:

\begin{observation}\label{obse:neighborhood}
For every vertex $v$ of $G$, $N(v)$ is the disjoint union of isolated vertices and at most one edge. Moreover, for every edge $xy \in E$, there is at most one common neighbor of $x$ and $y$.
\end{observation}

Since it is trivial to check whether $G$ has a d.i.m.\ $M$ with exactly one edge, from now on we can assume that $|M| \geq 2$. Since $G$ is connected and butterfly-free, we have:

\begin{observation}\label{obse:xy-in-P3}
If $|M| \geq 2$ then there is an edge in $M$ which is contained in a $P_3$ of $G$.
\end{observation}

\noindent
{\bf Proof.}
Let $xy \in M$ and assume that $xy$ is not part of an induced $P_3$ of $G$. Since $G$ is connected and $|M| \geq 2$, $(N(x) \cup N(y)) \setminus \{x,y\} \neq \emptyset$, and since we assume that  $xy$ is not part of an induced $P_3$ of $G$ and $G$ is $K_4$- and diamond-free, there is exactly one neighbor of $xy$, namely a common neighbor, say $z$ of $x$ and $y$. Again, since $|M| \geq 2$, $z$ has a neighbor $a \notin \{x,y\}$, and since $G$ is $K_4$- and diamond-free, $a,x,y,z$ induce a paw. Clearly, the edge $za$ is $xy$-excluded and has to be dominated by a second $M$-edge, say $ab \in M$ but now, since $G$ is butterfly-free, $zb \notin E$. Thus, $z,a,b$ induce a $P_3$ in $G$, and Observation \ref{obse:xy-in-P3} is shown.
\qed

\medskip

Recall \cite{BraMos2017} for Observation \ref{obse:xy-in-P3}.
Let $xy \in M$ be an $M$-edge for which there is a vertex $r$ such that $\{r,x,y\}$ induce a $P_3$ with edge $rx \in E$. By the assumption that $xy \in M$, we have that $x$ and $y$ are black, and it could lead to a feasible $xy$-coloring (if no contradiction arises).

\medskip

Let $N_0(xy):=\{x,y\}$ and for $i \ge 1$, let $$N_i(xy):=\{z \in V: dist_G(z,xy) = i\}$$ denote the {\em distance levels of $xy$}.
We consider a partition of $V$ into $N_i=N_i(xy)$, $i \ge 0$, with respect to the edge $xy$ (under the assumption that $xy \in M$).

Recall that by (\ref{IV(M)partition}), $V=V(M) \cup I $ is a partition of $V$ where $V(M)$ is the set of black vertices and
$I$ is the set of white vertices which is independent.

Since we assume that $xy \in M$ (and is an edge in a $P_3$), clearly, $N_1 \subseteq I$ and thus:
\begin{equation}\label{N1subI}
N_1 \mbox{ is an independent set of white vertices.}
\end{equation}

Moreover, no edge between $N_1$ and $N_2$ is in $M$. Since $N_1 \subseteq I$ and all neighbors of vertices in $I$ are in $V(M)$, we have:
\begin{equation}\label{N2M2S2}
N_2 \mbox{ is a set of black vertices and } G[N_2] \mbox{ is the disjoint union of edges and isolated vertices. }
\end{equation}

Let $M_2$ denote the set of edges $uv \in E$ with $u,v \in N_2$ and let $S_2 = \{u_1,\ldots,u_k\}$ denote the set of isolated vertices in $N_2$; $N_2=V(M_2) \cup S_2$ is a partition of $N_2$. Obviously:
\begin{equation}\label{M2subM}
M_2 \subseteq M \mbox{ and } S_2 \subseteq V(M).
\end{equation}

Obviously, by (\ref{M2subM}), we have:
\begin{equation}\label{M2xymandatory}
\mbox{Every edge in } M_2 \mbox{ is an $xy$-forced $M$-edge}.
\end{equation}

Thus, from now on, after applying the Edge Reduction for $M_2$-edges, we can assume that $V(M_2)=\emptyset$, i.e., $N_2=S_2 = \{u_1,\ldots,u_k\}$. For every $i \in \{1,\ldots,k\}$, let $u'_i \in N_3$ denote the {\em $M$-mate} of $u_i$ (i.e., $u_iu'_i \in M$). Let $M_3=\{u_iu'_i: 1 \le i \le k\}$ denote the set of $M$-edges with one endpoint in $S_2$ (and the other endpoint in $N_3$). Obviously, by (\ref{M2subM}) and the distance condition for a d.i.m.\ $M$, the following holds:
\begin{equation}\label{noMedgesN3N4}
\mbox{ No edge with both ends in } N_3 \mbox{ and no edge between } N_3 \mbox{ and } N_4 \mbox{ is in } M.
\end{equation}

As a consequence of (\ref{noMedgesN3N4}) and the fact that every triangle contains exactly one $M$-edge (recall Observation~\ref{dimC3C5C7C4} $(i)$), we have:
\begin{equation}\label{triangleaN3bcN4}
\mbox{For every $C_3$ $abc$} \mbox{ with } a \in N_3, \mbox{ and } b,c \in N_4, \mbox{ $bc \in M$ is an $xy$-forced $M$-edge}.
\end{equation}

This means that for the edge $bc$, the Edge Reduction can be applied, and from now on, we can assume that there is no such triangle $abc$ with $a \in N_3$ and $b,c \in N_4$, i.e., for every edge $uv \in E$ in $N_4$:
\begin{equation}\label{edgeN4N3neighb}
N(u) \cap N(v) \cap N_3 = \emptyset.
\end{equation}

\medskip

According to $(\ref{M2subM})$ and the assumption that $V(M_2)=\emptyset$ (recall $N_2 = \{u_1,\ldots,u_k\}$), let:
\begin{enumerate}
\item[ ] $T_{one} := \{t \in N_3: |N(t) \cap N_2| = 1\}$,

\item[ ] $T_i := T_{one} \cap N(u_i)$, $1 \le i \le k$, and

\item[ ] $S_3 := N_3 \setminus T_{one}$.
\end{enumerate}

By definition, $T_i$ is the set of {\em private} neighbors of $u_i \in N_2$ in $N_3$ (note that $u'_i \in T_i$),
$T_1 \cup \ldots \cup T_k$ is a partition of $T_{one}$, and $T_{one} \cup S_3$ is a partition of~$N_3$.

\begin{observation}[\cite{BraMos2017}]\label{lemm:structure2}
The following statements hold:
\begin{enumerate}
\item[$(i)$] For all $i \in \{1,\ldots,k\}$, $T_i \cap V(M)=\{u_i'\}$.
\item[$(ii)$] For all $i \in \{1,\ldots,k\}$, $T_i$ is the disjoint union of vertices and at most one edge.
\item[$(iii)$] $G[N_3]$ is bipartite.
\item[$(iv)$] $S_3 \subseteq I$, i.e., $S_3$ is an independent subset of white vertices.
\item[$(v)$] If a vertex $t_i \in T_i$ sees two vertices in $T_j$, $i \neq j$, $i,j \in \{1,\ldots,k\}$, then $u_it_i \in M$ is an $xy$-forced $M$-edge.
\end{enumerate}
\end{observation}

\noindent
{\bf Proof.} $(i)$: Holds by definition of $T_i$ and by the distance condition of a d.i.m.\ $M$.

\noindent
$(ii)$: Holds by Observation \ref{obse:neighborhood}.

\noindent
$(iii)$: Follows by Observation \ref{dimC3C5C7C4} $(i)$ since every odd cycle in $G$ must contain at least one $M$-edge, and by (\ref{noMedgesN3N4}).

\noindent
$(iv)$: If $v \in S_3:= N_3 \setminus T_{one}$, i.e., $v$ sees at least two $M$-vertices then clearly, $v \in I$, and thus, $S_3 \subseteq I$ is an independent subset (recall that $I$ is an independent set).

\noindent
$(v)$: Suppose that $t_1 \in T_1$ sees $a$ and $b$ in $T_2$. If $ab \in E$ then $u_2,a,b,t_1$ would induce a diamond in $G$. Thus, $ab \notin E$ and now,
$u_2,a,b,t_1$ induce a $C_4$ in $G$; by Observation \ref{dimC3C5C7C4} $(ii)$, no edge in the $C_4$ is in $M$, and by (\ref{noMedgesN3N4}), the only possible $M$-edge for dominating $t_1a,t_1b$ is $u_1t_1$, i.e., $t_1=u'_1$.
\qed

\medskip

By Observation \ref{lemm:structure2} $(iv)$ and the Vertex Reduction for the white vertices of $S_3$, we can assume:

\begin{itemize}
\item[(A1)] $S_3=\emptyset$, i.e., $N_3=T_1 \cup \ldots \cup T_k$.
\end{itemize}

By Observation \ref{lemm:structure2} $(v)$, we can assume:

\begin{itemize}
\item[(A2)] For $i,j \in \{1,\ldots,k\}$, $i \neq j$, every vertex $t_i \in T_i$ has at most one neighbor in $T_j$.
\end{itemize}

In particular, if for some $i \in \{1,\ldots,k\}$, $T_i=\emptyset$, then there is no d.i.m.\ $M$ of $G$ with $xy \in M$, and if $|T_i|=1$, say $T_i=\{t_i\}$, then $u_it_i$ is an $xy$-forced $M$-edge.
Thus, we can assume:

\begin{itemize}
\item[(A3)] For every $i \in \{1,\ldots,k\}$, $|T_i| \ge 2$.
\end{itemize}

Let us say that a vertex $t \in T_i$, $1 \le i \le k$, is an {\em out-vertex} of $T_i$ if it is adjacent to some vertex of $T_j$ with $j \neq i$,
or it is adjacent to some vertex of $N_4$, and $t$ is an {\em in-vertex} of $T_i$ otherwise.

For finding a d.i.m.\ $M$ with $xy \in M$, one can remove all but one in-vertices; that can be done in polynomial time.
In particular, if there is an edge between two in-vertices $t_1t_2 \in E$,  $t_1,t_2 \in T_i$, then either $t_1$ or $t_2$ is black, and thus, $T_i$ is completely colored. Thus, let us assume:

\begin{itemize}
\item[(A4)] For every $i \in \{1,\ldots,k\}$, $T_i$ has at most one in-vertex.
\end{itemize}

\begin{observation}\label{N4P6}
If $v \in N_i$ for $i \ge 4$ then $v$ is an endpoint of an induced $P_6$, say with vertices $v,v_1,v_2,v_3,v_4,v_5$ such that $v_1,v_2,v_3,v_4,v_5 \in \{x,y\} \cup N_1 \cup \ldots \cup N_{i-1}$ and with edges $vv_1 \in E$, $v_1v_2 \in E$, $v_2v_3 \in E$, $v_3v_4 \in E$, $v_4v_5 \in E$. Analogously, if $v \in N_3$ then $v$ is an endpoint of a corresponding induced $P_5$.
\end{observation}

\noindent
{\bf Proof.}
If $i \ge 5$ then clearly there is such a $P_6$. Thus, assume that $v \in N_4$. Then $v_1 \in N_3$ and $v_2 \in N_2$.
Recall that $y,x,r$ induce a $P_3$. If $v_2r \in E$ then $v,v_1,v_2,r,x,y$ induce a $P_6$. Thus assume that $v_2r \notin E$. Let $v_3 \in N_1$ be a
neighbor of $v_2$. Now, if $v_3x \in E$ then $v,v_1,v_2,v_3,x,r$ induce a $P_6$, and if  $v_3x \notin E$ but $v_3y \in E$ then $v,v_1,v_2,v_3,y,x$ induce a $P_6$.
Analogously, if $v \in N_3$ then $v$ is an endpoint of an induced $P_5$ (which could be part of the $P_6$ above).
Thus, Observation \ref{N4P6} is shown.
\qed

\begin{observation}\label{S115frN3S111fr}
There is no $S_{1,1,1}$ in $G[N_3]$ with vertices $a,b,c,d$ and center $a$ such that $d \in T_i$, $1 \le i \le k$, while no vertex of $\{a,b,c\}$
belongs to $T_i$.
\end{observation}

\noindent
{\bf Proof.} Suppose to the contrary that such an $S_{1,1,1}$ exists in $G[N_3]$. By Observation \ref{N4P6}, the vertex $d$ is the endpoint of a $P_5$ whose vertices are $d,u_i,v_1,v_2,v_3$ where $u_i$ is a neighbor of $d$ in $N_2$ and every vertex of $\{v_1,v_2,v_3\}$ belongs to $N_1 \cup \{x,y\}$. Since no vertex of $\{a,b,c\}$ is adjacent to a vertex of $\{u_i,v_1,v_2,v_3\}$, it follows that $\{a,b,c,d,u_i,v_1,v_2,v_3\}$ induces a $S_{1,1,5}$ in $G$ with center $a$, which is a contradiction.
\qed

\medskip

By Observation \ref{S115frN3S111fr}, we have:

\begin{observation}\label{N3TineighbTjatmost2}
Let $t_i$ be a vertex of $T_i$. The following statements hold:
\begin{enumerate}
\item[$(i)$] If $t_i \in T_i$ is part of an edge in $T_i$ and $t_i$ has a neighbor in $T_j$, $i \neq j$,
then $t_i$ has no other neighbor in $N_3 \setminus (T_i \cup T_j)$.
\item[$(ii)$] If $t_i \in T_i$ is not part of an edge in $T_i$ and $t_i$ has two neighbors in $T_j \cup T_s$ (possibly $j=s$), then $t_i$ has
no other neighbors in $N_3 \setminus (T_j \cup T_s)$.
\end{enumerate}
\end{observation}

\noindent
{\bf Proof.}
$(i)$: Without loss of generality, let $t_1t'_1 \in E$ for $t_1,t'_1 \in T_1$, and suppose to the contrary that $t_1$ has two neighbors $t_i \in T_i,t_j \in T_j$, $i \neq j$, $i,j \ge 2$.
Since $G$ is diamond-free, $t'_1t_i \notin E$ and $t'_1t_j \notin E$, and since $G$ is butterfly-free, $t_it_j \notin E$. But then $t_1,t'_1,t_i,t_j$ (with center $t_1$) induce an $S_{1,1,1}$ in $N_3$, which is a contradiction.

\noindent
$(ii)$: Now without loss of generality, $t_1 \in T_1$ is not part of an edge in $T_1$. Suppose to the contrary that $t_1$ has three neighbors $t_i \in T_i,t_j \in T_j,t_h \in T_h$ (possibly $i=j$) and $h \neq i,j$, then clearly, $t_i,t_j,t_h$ is independent (else there would be a triangle and thus, an $M$-edge in $N_3$ -- recall (\ref{noMedgesN3N4})). But then $t_1,t_i,t_j,t_h$ (with center $t_1$) induce an $S_{1,1,1}$ in $N_3$, which is a contradiction.
\qed

\begin{observation}\label{S115fr3edgesbetweenTiTj}
Assume that $G$ has a d.i.m.\ $M$ with $xy \in M$. Then there are no three edges between $T_i$ and $T_j$, $i \neq j$, and if there are two edges between $T_i$ and $T_j$, say $t_it_j \in E$ and $t'_it'_j \in E$ for $t_i,t'_i \in T_i$ and $t_j,t'_j \in T_j$ then any other vertex in $T_i$ or $T_j$ is white.
\end{observation}

\noindent
{\bf Proof.}
First, suppose to the contrary that there are three edges between $T_1$ and $T_2$, say $t_1t_2 \in E$, $t'_1t'_2 \in E$, and $t''_1t''_2 \in E$ for $t_i,t'_i,t''_i \in T_i$, $i=1,2$. Then $t_1$ is black if and only if $t_2$ is white, $t'_1$ is black if and only if $t'_2$ is white, and
$t''_1$ is black if and only if $t''_2$ is white. Without loss of generality, assume that $t_1$ is black, and $t_2$ is white. Then $t'_1$ is white, and $t'_2$ is black, but now, $t''_1$ and $t''_2$ are white, which is a contradiction.

Now, if there are exactly two such edges between $T_1$ and $T_2$, say $t_1t_2 \in E$, $t'_1t'_2 \in E$, then again, $t_1$ or $t'_1$ is black as well as $t_2$ or $t'_2$ is black, and thus, every other vertex in $T_1$ or $T_2$ is white.

\medskip

Thus Observation \ref{S115fr3edgesbetweenTiTj} is shown.
\qed

\medskip

By Observation \ref{S115fr3edgesbetweenTiTj}, we can assume:
\begin{itemize}
\item[(A5)] For $i,j \in \{1,\ldots,k\}$, $i \neq j$, there are at most two edges between $T_i$ and $T_j$.
\end{itemize}

Then let us introduce the following forcing rules (which are correct).
Since no edge in $N_3$ is in $M$ (recall (\ref{noMedgesN3N4})), we have:
\begin{itemize}
\item[(R1)] All $N_3$-neighbors of a black vertex in $N_3$ must be colored white, and all $N_3$-neighbors of a white vertex in
$N_3$ must be colored black.
\end{itemize}

Moreover, we have:
\begin{itemize}
\item[(R2)] Every $T_i$, $i \in \{1,\ldots,k\}$, should contain exactly one vertex which is black. Thus, if $t \in T_i$ is black then all the remaining vertices in $T_i \setminus \{t\}$ must be colored white.

\item[(R3)] If all but one vertices of $T_i$, $1 \le i \le k$, are white and the final vertex $t \in T_i$ is not yet colored, then $t$ must be colored black.
\end{itemize}

Since no edge between $N_3$ and $N_4$ is in $M$ (recall (\ref{noMedgesN3N4})), we have:
\begin{itemize}
\item[(R4)] For every edge $st \in E$ with $t \in N_3$ and $s \in N_4$, $s$ is white if and only if $t$ is black and vice versa.
\end{itemize}

Subsequently, for checking if $G$ has a d.i.m.\ $M$ with $xy \in M$, we consider the cases $N_4 = \emptyset$ and $N_4 \neq \emptyset$.

\section{The Case $N_4 = \emptyset$}\label{N4empty}

Let $A_{xy} := \{x,y\} \cup N_1 \cup N_2 \cup N_3$ and $B_{xy} := V \setminus A_{xy}$.
In this section we show that for the case $N_4=\emptyset$ (i.e., $B_{xy}=\emptyset$), one can check in polynomial time whether $G$ has a d.i.m.\ $M$ with $xy \in M$; we consider the feasible $xy$-colorings for $G[A_{xy}]$.
Recall that for every edge $uv \in M$, $u$ and $v$ are black, for $I=V(G) \setminus V(M)$, every vertex in $I$ is white, $N_2 = S_2 = \{u_1,\ldots,u_k\}$ and all $u_i$, $1 \le i \le k$, are black, $T_i=N(u_i) \cap N_3$, and recall assumptions (A1)-(A5) and rules (R1)-(R4). In particular, $S_3 = \emptyset$, i.e., $N_3=T_1 \cup \ldots \cup T_k$.

\medskip

Clearly, in the case $N_4 = \emptyset$, all the components of $G[S_2 \cup N_3]$ can be independently colored.
$G[\{u_i\} \cup T_i]$ is a {\em trivial component in $G[S_2 \cup N_3]$} if $T_i \cojoin T_j$ for every $j \neq i$.
Obviously, checking a possible d.i.m.\ $M$ with $xy \in M$ can be done easily (and independently) for trivial components; for a vertex $u'_i \in T_i$ let $u_iu'_i \in M$.

\medskip

From now on we can assume that every component $K$ in $G[S_2 \cup N_3]$ is nontrivial, i.e., $K$ contains at least two $T_i,T_j$, $i \neq j$ which contact to each other. Every component with at most four $S_2$-vertices has a polynomial number of feasible $xy$-colorings. Thus, we can focus on components with at least five $S_2$-vertices.

\begin{lemma}\label{S115frN3P5fr}
There is no $P_5$ $(t_1,t_2,t_3,t_4,t_5)$ in $G[N_3]$ with vertices $t_i \in T_i$, $1 \le i \le 5$.
\end{lemma}

\noindent
{\bf Proof.}
Suppose to the contrary that there is a $P_5$ $(t_1,t_2,t_3,t_4,t_5)$ in $G[N_3]$ with vertices $t_i \in T_i$, $1 \le i \le 5$.
Let $q_1 \in N_1$ be an $N_1$-neighbor of $u_1$, and without loss of generality, assume that $q_1x \in E$.

Since $q_1,x,u_5,u_1,t_1,t_2,t_3,t_4$ (with center $q_1$) do not induce an $S_{1,1,5}$ in $G$, we have $q_1u_5 \notin E$.

Since $q_1,x,u_1,u_3,t_3,t_4,t_5,u_5$ (with center $q_1$) do not induce an $S_{1,1,5}$ in $G$, we have $q_1u_3 \notin E$.

Since $t_4,t_5,u_4,t_3,t_2,t_1,u_1,q_1$ (with center $t_4$) do not induce an $S_{1,1,5}$ in $G$, we have $q_1u_4 \in E$.

But then $q_1,x,u_4,u_1,t_1,t_2,t_3,u_3$ (with center $q_1$) induce an $S_{1,1,5}$ in $G$, which is a contradiction.

\medskip

Thus, Lemma \ref{S115frN3P5fr} is shown.
\qed

\begin{lemma}\label{S115frN3P3}
If there is a $P_3$ $(t_1,t_2,t_3)$, $t_i \in T_i$, $1 \le i \le 3$, in a nontrivial component of $G[S_2 \cup N_3]$ then there is no other edge $t_it_j \in E$, $t_i \in T_i, t_j \in T_j$, $i,j \notin \{1,2,3\}$.
\end{lemma}

\noindent
{\bf Proof.}
Suppose to the contrary that there is another edge, say $t_4t_5 \in E$ for $t_4 \in T_4$ and $t_5 \in T_5$. Let $u_4 \in S_2$ with $u_4t_4 \in E$. Let $u_2 \in S_2$ with $u_2t_2 \in E$ and let $q_2 \in N_1$ be a neighbor of $u_2$. Clearly, $u_2t_i \notin E$ for $i \in \{1,3,4,5\}$ and
$u_4t_j \notin E$ for $j \in \{1,2,3,5\}$. If $u_2$ and $u_4$ do not have any common neighbor in $N_1$ then there would be an $S_{1,1,5}$ in $G[\{x,y\} \cup N_1 \cup \{u_2,u_4\}]$ with center $t_2$. Thus, assume that $q_2u_4 \in E$. If $\{t_1,t_2,t_3\} \cojoin \{t_4,t_5\}$ then $t_2,t_1,t_3,u_2,q_2,u_4,t_4,t_5$
(with center $t_2$) would induce an $S_{1,1,5}$. Thus, $\{t_4,t_5\}$ contacts $\{t_1,t_2,t_3\}$. Assume without loss of generality that $t_4$ is black. Then $t_5$ is white.

\medskip

\noindent
{\bf Case 1.} $t_2$ is black.

Then $t_1,t_3$ are white, and the only possible edges between $\{t_1,t_2,t_3\}$ and $\{t_4,t_5\}$ are $t_1t_4$, $t_3t_4$, or $t_2t_5$.

If $t_2t_5 \in E$ then $t_2,t_1,t_3,t_5$ (with center $t_2$) would induce an $S_{1,1,1}$. Thus, by Observation \ref{S115frN3S111fr}, $t_2t_5 \notin E$.

If $t_1t_4 \in E$ and $t_3t_4 \in E$ then $t_4,t_1,t_3,t_5$  (with center $t_4$) would induce an $S_{1,1,1}$. Thus, by Observation \ref{S115frN3S111fr},
either $t_1t_4 \notin E$ or $t_3t_4 \notin E$; without loss of generality, assume that $t_3t_4 \in E$ and $t_1t_4 \notin E$. But then $(t_1,t_2,t_3,t_4,t_5)$ induce a $P_5$ in $G[N_3]$, which is a contradiction to Lemma \ref{S115frN3P5fr}.

\medskip

\noindent
{\bf Case 2.} $t_2$ is white.

Then $t_1,t_3$ are black, and the only possible edges between $\{t_1,t_2,t_3\}$ and $\{t_4,t_5\}$ are $t_2t_4$, $t_1t_5$, or $t_3t_5$.

If $t_2t_4 \in E$ then $t_2,t_1,t_3,t_4$  (with center $t_2$) would induce an $S_{1,1,1}$. Thus, by Observation \ref{S115frN3S111fr}, $t_2t_4 \notin E$.

If $t_1t_5 \in E$ and $t_3t_5 \in E$ then $t_5,t_1,t_3,t_4$  (with center $t_5$) would induce an $S_{1,1,1}$. Thus, by Observation \ref{S115frN3S111fr},
either $t_1t_5 \notin E$ or $t_3t_5 \notin E$; without loss of generality, assume that $t_3t_5 \in E$ and $t_1t_5 \notin E$. But then $(t_1,t_2,t_3,t_5,t_4)$ induce a $P_5$ in $G[N_3]$, which is a contradiction to Lemma \ref{S115frN3P5fr}.

\medskip

Thus, Lemma \ref{S115frN3P3} is shown.
\qed

\begin{lemma}\label{S115frN3P3polcol}
If there is a $P_3$ $(t_1,t_2,t_3)$, $t_i \in T_i$, $1 \le i \le 3$, in a nontrivial component $K$ of $G[S_2 \cup N_3]$ then there are at most polynomially many feasible $xy$-colorings of $K$.
\end{lemma}

\noindent
{\bf Proof.}
Let $K=G[\{u_1,\ldots,u_p\} \cup T_1 \cup \ldots \cup T_p]$ be a nontrivial component in $G[S_2 \cup N_3]$, and without loss of generality, assume that $(t_1,t_2,t_3)$ is a $P_3$ with $t_i \in T_i$, $1 \le i \le 3$. By Lemma \ref{S115frN3P3}, any other $T_i$, $i \ge 4$, contacts $T_1 \cup T_2 \cup T_3$.
If $T_i$ contacts a white vertex $t \in T_1 \cup T_2 \cup T_3$, say $t_it \in E$ for $t_i \in T_i$, then $t_i$ is forced to be black, and thus,
$T_i$ is completely colored. Now we consider $T_i$ which only contacts a black vertex in $T_1 \cup T_2 \cup T_3$.

\medskip

\noindent
{\bf Case 1.} $t_2$ is black.

Then let $t'_1 \in T_1$ and $t'_3 \in T_3$ be black vertices. By Observations \ref{S115frN3S111fr} and \ref{N3TineighbTjatmost2}, each of $t'_1,t_2,t'_3$ contacts at most two $T_i$, $i \ge 4$. Thus, there are at most $n^6$ feasible $xy$-colorings of $K$. In this case, it can be shown even more, namely $t'_1$ as well as $t'_3$ have at most one such neighbor: Suppose to the contrary that $t'_3$ has two neighbors $t_i,t_j$. Then $t_i,t_j$ are white. Clearly, $t_it_j \notin E$, and $t_2t_i \notin E$, $t_2t_j \notin E$ (else there is an $S_{1,1,1}$ in $N_3$ with center $t_2$). Clearly, $t_3t'_3 \notin E$ (else there is an $S_{1,1,1}$ in $N_3$
with center $t'_3$). But now, $t'_3,t_i,t_j,u_3,t_3,t_2,t_1,u_1$ (with center $t'_3$) induce an $S_{1,1,5}$, which is a contradiction.
Thus, there are at most $n^2$ feasible $xy$-colorings of $K$.

\medskip

\noindent
{\bf Case 2.} $t_2$ is white.

Then $t_1,t_3$ are black, and let $t'_2 \in T_2$ be a black vertex.
Clearly, $t_1$ and $t_3$ have at most one neighbor in some $T_i$, and $t'_2$ has at most two such neighbors. Thus, there are at most $n^4$ feasible $xy$-colorings of $K$.

\medskip

Thus, Lemma \ref{S115frN3P3polcol} is shown.
\qed

\medskip

From now on we can assume:
\begin{itemize}
\item[(A6)] There is no $P_3$ $(t_i,t_j,t_h)$, $t_i \in T_i$, $t_j \in T_j$, $t_h \in T_h$,  in any nontrivial component $K$ of $G[S_2 \cup N_3]$.
\end{itemize}

We first consider cycles in $K$: Let $(t_1,(u_1),t'_1,t_2,(u_2),t'_2,\ldots,t_h,(u_h),t'_h)$ with $h \ge 2$ be a cycle in $G[S_2 \cup N_3]$; if $t_it'_i \in E$ then $u_i$ is not part of the cycle.

\begin{lemma}\label{S115frTicycles}
Assume that $G$ has a d.i.m.\ $M$ with $xy \in M$. Then for any cycle $(t_1,(u_1),$ $t'_1,t_2,(u_2),t'_2,\ldots,t_h,(u_h),t'_h)$ with $h \ge 2$ and $t_i,t'_i \in T_i$ such that $t'_it_{i+1} \in E$ $(i+1$ modulo $h)$, either $t_i$ or $t'_i$ is black, and thus, any other vertex in $T_i$ is white.
\end{lemma}

\noindent
{\bf Proof.}
Recall Observation \ref{S115fr3edgesbetweenTiTj} for $h=2$. Now assume that $h \ge 3$ and suppose to the contrary that $t_1$ and $t'_1$ are white. Then $t_2$ is black, $t'_2$ is white, $t_3$ is black and so on until finally, $t_h$ is black, $t'_h$ is white and the edge $t'_ht_1$ consists of two white vertices, which is a contradiction. Thus, either $t_i$ or $t'_i$ is black, and thus, any other vertex in $T_i$ is white.

\medskip

Thus, Lemma \ref{S115frTicycles} is shown.
\qed

\medskip

Every such cycle in $K$ has at most two feasible $xy$-colorings. Now, for coloring $K$, we start with a cycle $C$ in $K$ and one of the two feasible $xy$-colorings of $C$. Then every out-vertex $t''_i \in T_i$ having a neighbor in some $T_j$ which is not part of $C$ is forced to be white since either $t_i$ or $t'_i$ (which are part of the cycle $C$) are black (and by (A6), the black vertex of $t_i,t'_i$ has no other neighbor in some $T_j$). This leads to a black-forced neighbor $t_j \in T_j$, $h+1 \le j \le p$, of $t''_i$, and it continues in the same way with every other contact in $K$. For the second feasible $xy$-coloring of $C$, it will be done in the same way.

Thus, the DIM problem for $K$ (i.e., there is either a d.i.m.\ for $K$ or there is a contradiction) is solved in polynomial time.

\medskip

If $K$ has no chordless cycle then DIM for $K$ can be solved in polynomial time since in this case, $K$ is $K_4$-free chordal and thus, the treewidth and the clique-width of $K$ is bounded.
In \cite{CarKorLoz2011}, it was mentioned that DIM is solvable in polynomial time for graph classes with bounded clique-width.

\medskip

Summarizing, in the case $N_4 = \emptyset$, the previous results show that DIM for $K$ has a polynomial time solution since all the components of $G[S_2 \cup N_3]$ can be independently colored. This leads to:

\begin{theorem}
If $N_4 = \emptyset$ then one can check in polynomial time whether $G$ has a d.i.m.\ containing $xy$. \qed
\end{theorem}

\section{The Case $N_4 \neq \emptyset$}\label{N4nonempty}

Recall $A_{xy} := \{x,y\} \cup N_1 \cup N_2 \cup N_3$ and $B_{xy} := V \setminus A_{xy}$.

\medskip

Next we show:
\begin{lemma}\label{GBxyS111fr}
$G[B_{xy} \setminus N_4]$ is $S_{1,1,1}$-free.
\end{lemma}

\noindent
{\bf Proof.}
Suppose to the contrary that there is an $S_{1,1,1}$ in $G[B_{xy} \setminus N_4]$, say with vertices $a,b,c,d$ and edges $ab,ac,ad \in E$ (i.e., center $a$). Assume that the minimum distance level containing a vertex of $a,b,c,d$ is $N_p$, $p \ge 5$, and let $z \in N_{p-1}$ be a neighbor of the $S_{1,1,1}$ and $(z,z_2,\ldots,z_6)$ be a $P_6$ in $G[\{x,y\} \cup N_1 \cup \ldots \cup N_{p-1}]$ as in Observation \ref{N4P6}.

First assume that $za \in E$. Since $z,a,b,c$ do not induce a diamond, we have $zb \notin E$ or $zc \notin E$, and analogously, $z,a,b,d$ as well as $z,a,c,d$ do not induce a diamond. Then $z$ is adjacent to at most one of $b,c,d$, say without loss of generality, $zb \in E$, $zc \notin E$, and $zd \notin E$. But then
$a,c,d,z,z_2,\ldots,z_5$ (with center $a$) induce an $S_{1,1,5}$, which is a contradiction. Thus,  $za \notin E$.

Now, without loss of generality, assume that $zb \in E$. Since $z,b,c,z_2,\ldots,z_6$ (with center $z$) do not induce an $S_{1,1,5}$, we have $zc \notin E$, and analogously, since $z,b,d,z_2,\ldots,z_6$ (with center $z$) do not induce an $S_{1,1,5}$, we have $zd \notin E$. But now,
$a,c,d,b,z,z_2,z_3,z_4$ (with center $a$) induce an $S_{1,1,5}$, which is a contradiction.

\medskip

Thus, Lemma \ref{GBxyS111fr} is shown.
\qed

\subsection{The Case $|N_2| \le 4$}\label{BxynonemptyN2le3}

\begin{lemma}\label{prop:coloringG[X]}
If $|N_2| \le 4$, then $G[A_{xy}]$ has at most $O(n^4)$ feasible $xy$-colorings.
\end{lemma}

\noindent
{\bf Proof.} This directly follows by the above structure properties since $|N_2| \leq 4$ say $N_3 = T_1 \cup T_2 \cup T_3 \cup T_4$. In particular, for each $(t_1,t_2,t_3,t_4)$ $\in$ $T_1 \times T_2 \times T_3 \times T_4$, one can assign the color black to vertices $t_1,t_2,t_3,t_4$ (and then assign the color white to the remaining vertices of $T_1 \cup T_2 \cup T_3 \cup T_4$), and check if $\{xy,u_1t_1,u_2t_2,u_3t_3,u_4t_4\}$ is a d.i.m.\ of $G[A_{xy}]$.
\qed

\begin{lemma}\label{prop:N4}
If the colors of all vertices in $G[A_{xy}]$ are fixed then the colors of all vertices in $N_4$ are forced.
\end{lemma}

\noindent
{\bf Proof.} Let $v \in N_4$ and let $w \in N_3$ be a neighbor of $v$. Since by (\ref{noMedgesN3N4}), every edge between $N_3$ and $N_4$ is $xy$-excluded, we have:
If $w$ is white then $v$ is black, and if $w$ is black then $v$ is white.
\qed

\begin{theorem}\label{theo:N2atmost4}
If $|N_2| \leq 4$ then one can check in polynomial time whether $G$ has a d.i.m.\ containing $xy$.
\end{theorem}

\noindent
{\bf Proof.} The proof is given by the following procedure:

\begin{proc}\label{DIMwithxyCase1}

\begin{tabbing}	
xxxxxxx \= \kill\\
{\bf Input:} \> A connected $(S_{1,1,5})$-free graph $G = (V,E)$, and\\
\> an edge $xy \in E$, which is part of a $P_3$ in $G$.\\
{\bf Task:} \> Return either a d.i.m.\ $M$ with $xy \in M$\\
\> or a proof that $G$ has no d.i.m.\ containing $xy$.
\end{tabbing}

\begin{itemize}
\item[$(a)$]  {\bf if} $N_4 = \emptyset$ {\bf then} apply the approach described in Section $\ref{N4empty}$. Then return either a d.i.m.\ $M$ with $xy \in M$ or
``$G$ has no d.i.m.\ containing $xy$''.

\item[$(b)$]  {\bf if} $N_4 \neq \emptyset$ {\bf then} for $A_{xy} := \{x,y\} \cup N_1 \cup N_2 \cup N_3$ and $B_{xy} := V \setminus A_{xy}$ {\bf do}
\begin{itemize}
\item[$(b.1)$] Compute all black-white $xy$-colorings of $G[A_{xy}]$ by Lemma $\ref{prop:coloringG[X]}$. If no such $xy$-coloring without contradiction exists, then return ``$G$ has no d.i.m.\ containing $xy$''

\item[$(b.2)$]  {\bf for each} $xy$-coloring of $G[A_{xy}]$  {\bf do}

\begin{enumerate}

\item[ ]  Derive a partial $xy$-coloring of $G[B_{xy}]$ by the forcing rules; in particular all vertices of $N_4$ will be colored by Lemma $\ref{prop:N4}$ according to the $xy$-coloring of $G[A_{xy}]$;
\item[ ] {\bf if} a contradiction arises in vertex coloring {\bf then} proceed to the next $xy$-coloring of $G[A_{xy}]$
\item[ ] {\bf else} apply the algorithm of Cardoso et al.\ in $\cite{CarKorLoz2011}$  $($see Theorem $\ref{DIMpolresults}$ $(i))$ to determine if $G[B_{xy} \setminus N_4]$ (which is claw-free by Lemma $\ref{GBxyS111fr}$) has a d.i.m.
\item[ ] {\bf if} $G[B_{xy} \setminus N_4]$ has a d.i.m.\ {\bf then} STOP and return the $xy$-coloring of $G$ derived by the $xy$-coloring of $G[X]$ and by such a d.i.m.\ of $G[B_{xy}]$.

\end{enumerate}

\item[$(b.3)$] STOP and return ``$G$ has no d.i.m.\ containing $xy$''.
\end{itemize}
\end{itemize}

\end{proc}

The correctness of Procedure \ref{DIMwithxyCase1} follows from the structural analysis of $S_{1,1,5}$-free graphs with a d.i.m.\ and by the results in the present section.

\medskip

The polynomial time bound of Procedure \ref{DIMwithxyCase1} follows from the fact that
Step (a) can be done in polynomial time by the results in Section \ref{N4empty}, and
Step (b) can be done in polynomial time by Lemma \ref{prop:coloringG[X]}, since the forcing rules can be executed in polynomial time, and since the solution algorithm of Cardoso et al.\ (see Theorem \ref{DIMpolresults} $(i)$) can be executed in polynomial time. Thus, Theorem \ref{theo:N2atmost4} is shown.
\qed

\subsection{The Case $|N_2| \ge 5$}\label{N4nonemptyN2ge5}

\begin{lemma}\label{N2ge4uicommonneighb}
If $|N_2| \ge 5$ and there is an end-vertex $u_i \in N_2$ of a $P_4$ $(u_i,t_i,z_1,z_2)$ with $t_i \in T_i$, $z_1 \in N_4$, and $z_2 \in N_4 \cup N_5$
then there exists a vertex $q_i \in N_1$ such that $q_iu_i \in E$ and $q_i$ has a second neighbor $u_j \in N_2$, $i \neq j$, i.e., $u_i$ and $u_j$ have a common $N_1$-neighbor.
\end{lemma}

\noindent
{\bf Proof.}
Without loss of generality, assume that $(u_1,t_1,z_1,z_2)$ is a $P_4$ with $u_1 \in S_2$, $t_1 \in T_1$, $z_1 \in N_4$, and $z_2 \in N_4 \cup N_5$. Let $q_1 \in N_1$ be a neighbor of $u_1$. Recall that by Observation~\ref{obse:neighborhood}, $x$ and $y$ have at most one common neighbor in $N_1$, and recall that $|N_2| \ge 5$ in this section.

\medskip

First assume that $q_1$ is a common neighbor of $x$ and $y$.  Then there are at least two vertices $u_2,u_3 \in N_2$ with $N_1$-neighbors $q_2,q_3$ which see either the same $x$ or $y$ (but not both of them); without loss of generality, assume that $q_2u_2 \in E$ and $q_3u_3 \in E$ for $q_2x \in E$ and $q_3x \in E$. Suppose to the contrary that $u_2,u_3$ do not see $q_1$ (else $u_1$ has a common $N_1$-neighbor with $u_2$ or $u_3$).

If $u_3q_2 \in E$ then, since $(q_2,u_2,u_3,x,q_1,u_1,t_1,z_1)$ do not induce an $S_{1,1,5}$ with center $q_2$, we have $u_1q_2 \in E$, i.e., $u_1$ and $u_2$ have a common $N_1$-neighbor $q_2$.

If $u_3q_2 \notin E$ and $u_3q_3 \in E$, $q_2u_2 \in E$, $u_2q_3 \notin E$ with $q_2x \in E$ and $q_3x \in E$ then, since $(x,q_2,q_3,q_1,u_1,t_1,z_1,z_2)$ do not induce an $S_{1,1,5}$ with center $x$, we have $u_1q_2 \in E$ or $u_1q_3 \in E$, i.e., $u_1$ and $u_2$ have a common $N_1$-neighbor $q_2$ or $u_1$ and $u_3$ have a common $N_1$-neighbor $q_3$.

\medskip

Next assume that $q_1$ is no common neighbor of $x$ and $y$, say without loss of generality, $q_1x \in E$ and $q_1y \notin E$.
Suppose to the contrary that $u_2$ does not see $q_1$ (else $u_1$ has a common $N_1$-neighbor with $u_2$). Let $q_2 \in N_1$ with $q_2u_2 \in E$. If $q_2x \in E$ and $q_2y \notin E$ then, since $(x,q_2,y,q_1,u_1,t_1,z_1,z_2)$ do not induce an $S_{1,1,5}$ with center $q_2$, we have $u_1q_2 \in E$, i.e., $u_1$ and $u_2$ have a common $N_1$-neighbor $q_2$.

\medskip

Now assume that $q_2y \in E$. First assume that $q_2x \in E$ and $q_2y \in E$. If there are two neighbors $u_2,u_3 \in N_2$ of $q_2$ which do not see $q_1$ then, since $(q_2,u_2,u_3,x,q_1,u_1,t_1,z_1)$ do not induce an $S_{1,1,5}$ with center $q_2$, we have $u_1q_2 \in E$, i.e., $u_1$ and $u_2$ have a common $N_1$-neighbor $q_2$. Now assume that $q_2$ has only one $N_2$-neighbor $u_2$. Clearly, every other vertex $u_i, i \ge 3$, in $N_2$ has an $N_1$-neighbor which sees only $y$.

Assume that $q_3u_3 \in E$ and $q_3u_4 \in E$ as well as $q_3y \in E$, and suppose to the contrary that $u_3,u_4$ do not see $q_1$. Then, since
 $(q_3,u_3,u_4,y,x,q_1,u_1,t_1)$ do not induce an $S_{1,1,5}$ with center $q_3$, we have $u_1q_3 \in E$, i.e., $u_1$ and $u_3$ have a common $N_1$-neighbor $q_3$.

Finally assume that $q_3u_3 \in E$ and $q_3u_4 \notin E$, $q_4u_4 \in E$ and $q_4u_3 \notin E$, as well as $q_3y \in E$ and $q_4y \in E$.
Then, since $(y,q_3,q_4,x,q_1,u_1,t_1,z_1)$ do not induce an $S_{1,1,5}$ with center $q_3$, we have $u_1q_3 \in E$ or $u_1q_4 \in E$, i.e., $u_1$ and $u_3$ have a common $N_1$-neighbor $q_3$, or $u_1$ and $u_4$ have a common $N_1$-neighbor $q_4$.

\medskip

Thus, Lemma \ref{N2ge4uicommonneighb} is shown.
\qed

\begin{lemma}\label{N2atleast4noP5endui}
If $|N_2| \ge 5$ then there is no $P_5$ $(u_i,t_i,z_1,z_2,z_3)$ with $u_i \in N_2$, $t_i \in T_i$, $z_1 \in N_4$, $z_2 \in N_4 \cup N_5$, and $z_3 \in N_4 \cup N_5 \cup N_6$.
\end{lemma}

\noindent
{\bf Proof.}
Suppose to the contrary that there is such a $P_5$ $(u_1,t_1,z_1,z_2,z_3)$ with $u_1 \in N_2$, $t_1 \in T_1$, $z_1 \in N_4$, $z_2 \in N_4 \cup N_5$, and
$z_3 \in N_4 \cup N_5 \cup N_6$. Let $q_1 \in N_1$ with $q_1u_1 \in E$, and without loss of generality, let $q_1x \in E$.
By Lemma \ref{N2ge4uicommonneighb}, there is a second vertex $u' \in N_2$ with $q_1u' \in E$.
But then $q_1,x,u',u_1,t_1,z_1,z_2,z_3$ (with center $q_1$) induce an $S_{1,1,5}$, which is a contradiction.

\medskip

Thus, Lemma \ref{N2atleast4noP5endui} is shown.
\qed

\medskip

Since $N_6 \neq \emptyset$ would lead to such a $P_5$ as in Lemma \ref{N2atleast4noP5endui}, we have:

\begin{lemma}\label{S115frN2atleast4N6empty}
If $|N_2| \ge 5$ then $N_6 = \emptyset$.
\end{lemma}

Thus, from now on, we can assume that $B_{xy}=N_4 \cup N_5$.

\subsubsection{The Case $N_5 = \emptyset$}\label{N5empty}

In this case, we show that one can check in polynomial time whether $G$ has a d.i.m.\ $M$ with $xy \in M$.
Recall Lemma \ref{prop:N4}.

\medskip

Let $K$ be a nontrivial component of $G[S_2 \cup N_3 \cup N_4]$. Clearly, $K$ can have several components in $G[S_2 \cup N_3]$ which are connected by some $N_4$-vertices. $K$ can be colored by starting with a component in $G[S_2 \cup N_3]$ which is part of $K$.

\medskip

For every $i \in \{1,\ldots,k\}$, let $$Ext(T_i) := N(T_i) \cap N_4.$$

Since $N_5 = \emptyset$ and by (\ref{noMedgesN3N4}), every edge between $N_3$ and $N_4$ is $xy$-excluded, we have:
\begin{equation}\label{N4isolatedvertexwhite}
\mbox{If } v \in N_4 \mbox{ with } N(v) \cap N_4=\emptyset \mbox{ then } v \mbox{ is white}.
\end{equation}

Thus, from now on, by the Vertex Reduction, we can assume that every vertex in $N_4$ has a neighbor in $N_4$. Recall that $G$ is (diamond,butterfly)-free and therefore triangles are disjoint.
We first claim:

\begin{lemma}\label{N5emptyN4P3fr}
$G[N_4]$ is $P_3$-free.
\end{lemma}

\noindent
{\bf Proof.}
Suppose to the contrary that there is a $P_3$ $(a,b,c)$ in $G[N_4]$, and let $t_1 \in T_1$ be a neighbor of $b$ in $N_3$. Recall that
by (\ref{edgeN4N3neighb}), $t_1a \notin E$ and $t_1c \notin E$ (and recall Observation \ref{N4P6} where $t_1$ is the endpoint of a $P_5$). But then, $b,a,c,t_1,u_1,q_1,x,y$ (with center $b$) induce an $S_{1,1,5}$ if for the $N_1$-neighbor $q_1$ of $u_1$, $q_1x \in E$ and $q_1y \notin E$, and correspondingly for the other cases of $q_1$-neighbors (since $y,x,r$ induce a $P_3$).

\medskip

Thus, Lemma \ref{N5emptyN4P3fr} is shown.
\qed

\medskip

Now let us consider an edge $vw \in E$, with $v,w \in N_4$, which is isolated in $G[N_4]$.
Again, since $N_5 = \emptyset$ and by (\ref{noMedgesN3N4}), every edge between $N_3$ and $N_4$ is $xy$-excluded, we have: If $v$ is white then $w$ is black but there exists no vertex $w'$ such that $ww' \in M$, i.e., there is no d.i.m.\ $M$ with $xy \in M$. Thus:
\begin{equation}\label{N4isolatededgeblack}
\mbox{If for } v,w \in N_4, N(v) \cap N_4 =\{w\} \mbox{ and } N(w) \cap N_4 =\{v\} \mbox{ then } v,w \mbox{ are black}.
\end{equation}

Next we show:

\begin{lemma}\label{N5emptyN4Extu1C3}
At most one $Ext(T_i)$ has a $C_3$ in $G[N_4]$.
\end{lemma}

\noindent
{\bf Proof.}
Suppose to the contrary that there are two such cases. Without loss of generality, suppose that there is a $C_3$ $z_1z_2z_3$ in $Ext(T_1)$, and there is a $C_3$ $z_4z_5z_6$ in $Ext(T_2)$. Without loss of generality, assume that $z_1,z_2$ are black and thus, $z_3$ is white as well as $z_4,z_5$ are black and thus, $z_6$ is white. Let $t_1,t'_1,t''_1 \in T_1$ be the neighbors of $z_1,z_2,z_3$, i.e., $t_1z_1 \in E$, $t'_1z_2 \in E$, and $t''_1z_3 \in E$ as well as
$t_2z_4 \in E$, $t'_2z_5 \in E$, and $t''_2z_6 \in E$ for $t_2,t'_2,t''_2 \in T_2$. Recall that by (\ref{edgeN4N3neighb}), $t_1,t'_1,t''_1 \in T_1$ are distinct, and
 $t_2,t'_2,t''_2 \in T_2$ are distinct. By the colors of $z_1,\ldots,z_6$, we have that $t_1,t'_1,t_2,t'_2$ are white and $t''_1,t''_2$ are black.

First assume that $u_1$ and $u_2$ do not have a common neighbor in $N_1$; for $q_1,q_2 \in N_1$, let $q_1u_1 \in E$ and $q_2u_2 \in E$ but $q_1u_2 \notin E$ and $q_2u_1 \notin E$. Recall that $t_1,t'_1,t_2$ are white.
If $q_1x \in E$ and $q_2x \in E$ then $u_1,t_1,t'_1,q_1,x,q_2,u_2,t_2$ (with center $u_1$) would induce an $S_{1,1,5}$, and analogously, for $q_1y \in E$ and $q_2y \in E$. If $q_1$ and $q_2$ do not have a common neighbor $x$ or $y$ then without loss of generality, let $q_1x \in E$, $q_1y \notin E$, and $q_2y \in E$, $q_2x \notin E$. But then $u_1,t_1,t'_1,q_1,x,y,q_2,u_2$ (with center $u_1$) induce an $S_{1,1,5}$, which is a contradiction.

\medskip

Thus, we can assume that there is a common neighbor $q_1 \in N_1$, $q_1u_1 \in E$, $q_1u_2 \in E$.
By the colors (recall that $t_2,t'_2,t_1,z_3$ are white), and since $t_2,t'_2,u_2,q_1,u_1,t_1,z_1,z_3$ (with center $u_2$) do not induce an $S_{1,1,5}$, we have $t_2z_1 \in E$ or $t'_2z_1 \in E$; without loss of generality, assume that $t_2z_1 \in E$.
Recall Observation \ref{N4P6} where $t_2$ is the endpoint of a $P_5$.

But then $z_1,t_1,z_3,t_2$ and the $P_5$ with endpoint $t_2$ induce an $S_{1,1,5}$ (with center $z_1$), which is a contradiction.

\medskip

Thus, Lemma \ref{N5emptyN4Extu1C3} is shown.
\qed

\medskip

If there is no $C_3$ in any $Ext(T_i)$ then clearly, by Lemma \ref{N5emptyN4P3fr}, (\ref{N4isolatedvertexwhite}) and (\ref{N4isolatededgeblack}), the colors of all vertices in $K$ are forced. If there is a $C_3$ in exactly one of the $Ext(T_i)$, then for the three possible $M$-edges in the $C_3$, one can color all the other vertices accordingly in polynomial time. This leads to:

\begin{theorem}\label{S115frxydim}
If $|N_2| \geq 5$ and $N_5 = \emptyset$ then one can check in polynomial time whether $G$ has a d.i.m.\ containing $xy$.
\end{theorem}

\subsubsection{The Case $N_5 \neq \emptyset$}\label{N5nonempty}

\begin{lemma}\label{lemm:structure4}
Let $z \in N_4$. Then the following statements hold:
\begin{enumerate}
\item[$(i)$] If $z$ has a neighbor $z' \in N_5$ and $z'$ has a neighbor $z'' \in N_5$ then $zz'' \in E$.
\item[$(ii)$] If $z$ has a neighbor in $N_5$ then $N(z) \cap N_5$ is an edge or a single vertex.
\item[$(iii)$] If $z$ contacts $T_i$ and $T_j$ for $i \neq j$ then $N(z) \cap N_5 = \emptyset$.
\end{enumerate}
\end{lemma}

\noindent
{\bf Proof.}
$(i)$: Let $t_i \in T_i$ be a neighbor of $z$. By Lemma \ref{N2atleast4noP5endui}, $(u_i,t_i,z,z',z'')$ is no $P_5$. Thus, we have $zz'' \in E$.

\noindent
$(ii)$: If $z$ has two independent neighbors $w_1,w_2 \in N(z) \cap N_5$, i.e., $w_1w_2 \notin E$ then by Observation~\ref{N4P6}, $z$ is an endpoint of a $P_6$ in $\{x,y\} \cup N_1 \cup \ldots \cup N_4$, and it leads to an $S_{1,1,5}$ with the $P_6$ and $w_1,w_2$ (with center $z$), which is a contradiction. Thus, $N(z) \cap N_5$ is an edge or a single vertex.

\noindent
$(iii)$: Let $z$ contact $T_i$ and $T_j$ for $i \neq j$, say $zt_i \in E$ and $zt_j \in E$ for $t_i \in T_i$ and $t_j \in T_j$. By  (\ref{noMedgesN3N4}), $t_it_j \notin E$ (else, $z,t_i,t_j$ induce a triangle, and then, there is no d.i.m.\ $M$ with $xy \in M$).

Suppose to the contrary that $z$ contacts $N_5$, say $zz' \in E$ for some $z' \in N_5$. But then, by Observation~\ref{N4P6}, it leads to an $S_{1,1,5}$ with the $P_6$ along $t_i$, and with $t_j,z'$ (with center $z$).

\medskip

Thus, Lemma \ref{lemm:structure4} is shown.
\qed

\begin{lemma}\label{N5P3C3fr}
$G[N_5]$ is $P_3$- and $C_3$-free.
\end{lemma}

\noindent
{\bf Proof.}
First suppose to the contrary that there is a $P_3$ $(a,b,c)$ in $G[N_5]$. Let $z_a \in N_4$ be a neighbor of $a$. Then, by Lemma \ref{lemm:structure4} $(ii)$,
$z_ac \notin E$. Let $t_i \in T_i$ be a neighbor of $z_a$. By Lemma~\ref{N2atleast4noP5endui}, $(b,a,z_a,t_i,u_i)$ do not induce a $P_5$. Thus, $z_ab \in E$ but now
$(c,b,z_a,t_i,u_i)$ induce a $P_5$,  which is a contradiction. Thus, $G[N_5]$ is $P_3$-free.

Now suppose to the contrary that there is a $C_3$ $(a,b,c)$ in $G[N_5]$. Let again $z_a \in N_4$ be a neighbor of $a$. Since $G$ is $(K_4$,diamond)-free, $z_a$ is
nonadjacent to $b$ and $c$, and let again $t_i \in T_i$ be a neighbor of $z_a$ but now, $(b,a,z_a,t_i,u_i)$ induce a $P_5$, which is a contradiction. Thus, $G[N_5]$ is $C_3$-free.

\medskip

Thus, Lemma \ref{N5P3C3fr} is shown.
\qed

\begin{lemma}\label{N4N5P3fr}
There is no $P_3$ in $G[N_4 \cup N_5]$ with at least one $N_5$-end-vertex.
\end{lemma}

\noindent
{\bf Proof.}
By Lemma \ref{N5P3C3fr}, there is no $P_3$ $(a,b,c)$ with $a,b,c \in N_5$.
By Lemma \ref{lemm:structure4} $(i)$, there is no $P_3$ $(a,b,c)$ with $a \in N_4$ and $b,c \in N_5$.
By Lemma \ref{lemm:structure4} $(ii)$, there is no $P_3$ $(a,b,c)$ with $b \in N_4$ and $a,c \in N_5$.

Finally, let $(a,b,c)$ be a $P_3$ with $a,b \in N_4$ and $c \in N_5$. Since for a neighbor $t_i \in N_3$ of vertex $a$, $(u_i,t_i,a,b,c)$ would induce a $P_5$,
we have $t_ib \in E$, i.e., $t_i,a,b$ induce a $C_3$, but by (\ref{edgeN4N3neighb}), after the Edge Reduction, there is no such triangle.

\medskip

Thus, Lemma \ref{N4N5P3fr} is shown.
\qed

\medskip

Recall that by $(\ref{noMedgesN3N4})$, every edge between $N_3$ and $N_4$ is $xy$-excluded (and recall (\ref{N4isolatedvertexwhite})). Thus we have:
\begin{equation}\label{N4isolated}
\mbox{If } v \in N_4 \mbox{ is isolated in } N_4 \cup N_5 \mbox{ then } v \mbox{ is white}.
\end{equation}

Now, by the Vertex Reduction, we can assume that every vertex in $N_4$ has a neighbor in $N_4 \cup N_5$.

\begin{lemma}\label{N5isolatedwhite}
If $z \in N_5$ is isolated in $G[N_5]$ and $z$ has at least two nonadjacent neighbors in $N_4$ then $z$ is $xy$-forced to be white.
\end{lemma}

\noindent
{\bf Proof.}
Let $z_1,z_2 \in N_4$ be neighbors of $z \in N_5$, $z_1z_2 \notin E$, and let $t_i \in N_3$ be a neighbor of $z_1$. By Lemma \ref{N2atleast4noP5endui},
$(u_i,t_i,z_1,z,z_2)$ do not induce a $P_5$. Thus, $t_i$ is a common neighbor of $z_1$ and $z_2$. Suppose to the contrary that $z$ is black. Then, since $t_i,z_1,z_2,z$ induce a $C_4$, $t_i$ is black.
Since $z \in N_5$ is isolated in $G[N_5]$, there is a black neighbor $z' \in N_4$ of $z$ (if there is such a d.i.m.\ $M$ in $G$ with $xy \in M$). Since $G$ is diamond-free, $z'$ does not see both of $z_1,z_2$, say $z'z_1 \notin E$. But then $(u_i,t_i,z_1,z,z')$ induce a $P_5$, which is a contradiction to Lemma \ref{N2atleast4noP5endui}.

\medskip

Thus, Lemma \ref{N5isolatedwhite} is shown.
\qed

\medskip

Now, by the Vertex Reduction, we can assume that every vertex which is isolated in $G[N_5]$ has either exactly one neighbor or exactly two adjacent neighbors
in $N_4$.

\medskip

A component of $G[N_4 \cup N_5]$ with at least one vertex in $N_5$ is {\em trivial} if it has exactly two vertices, namely $a \in N_4$ and $b \in N_5$ with $ab \in E$.

\begin{lemma}\label{compN4N5}
Each nontrivial component of $G[N_4 \cup N_5]$ with at least one vertex in $N_5$, is a triangle $abc$ with one of the following two types:
\begin{enumerate}
\item[$(i)$] $a,b \in N_4$, and $c \in N_5$, or
\item[$(ii)$] $a \in N_4$, and $b,c \in N_5$.
\end{enumerate}
\end{lemma}

\noindent
{\bf Proof.}
Let $K$ be a nontrivial component of $G[N_4 \cup N_5]$ with at least one vertex in $N_5$.

First assume that $K$ has a triangle $abc$ with $a,b \in N_4$, and $c \in N_5$.
By Lemma \ref{N4N5P3fr}, there is no $P_3$ with at least one end-vertex in $N_5$.

If $c$ has a neighbor $d \in N_5$ then, since $(a,c,d)$ do not induce a $P_3$, we have $ad \in E$, and since $(b,c,d)$ do not induce a $P_3$,
we have $bd \in E$, but then $a,b,c,d$ induce a $K_4$ in $G$, which is a contradiction. Thus, vertex $c$ is isolated in $N_5$.

If one of $a,b$ has another neighbor in $N_4$, say $bd \in E$ for a vertex $d \in N_4$ then, since $(d,b,c)$ do not induce a $P_3$, we have
$cd \in E$, which is a contradiction since $G$ is (diamond,$K_4$)-free.

\medskip

Next assume that $K$ has a triangle $abc$ with $a \in N_4$, and $b,c \in N_5$.
By the same arguments, since there is no $P_3$ with at least one end-vertex in $N_5$, and since $G$ is (diamond,$K_4$)-free, there is no other neighbor of
$abc$ in $N_4 \cup N_5$.

\medskip

Thus, Lemma \ref{compN4N5} is shown.
\qed

\begin{lemma}\label{N4colorimplyN5color}
If the colors of all vertices in $N_4$ are fixed then it implies the colors of all vertices in $N_5$.
\end{lemma}

\noindent
{\bf Proof.}
Let $K$ be a component in $G[N_4 \cup N_5]$. First assume that $K$ is trivial with $ab \in E$ for $a \in N_4$ and $b \in N_5$. If $a$ is white then
$G$ has no d.i.m.\ $M$ with $xy \in M$. If $a$ is black then, since the edges between $N_3$ and $N_4$ are $xy$-excluded, and $K$ is trivial (i.e.,
 $a$ is nonadjacent to any black vertex in $N_4$), $b$ is black.

Now assume that $K$ is nontrivial of type $(i)$ in Lemma \ref{compN4N5}. If $a$ or $b$ is white then $c$ is black, else $c$ is white.

Finally, assume that $K$ is nontrivial of type $(ii)$ in Lemma \ref{compN4N5}. If $a$ is white then $b$ and $c$ are black. If $a$ is black then by Lemma \ref{compN4N5} $(ii)$, $a$ is nonadjacent to any black vertex in $N_4$, and then the $M$-mate of $a$ is $b$ or $c$, i.e., $b$ is black if and only if $c$ is white.

\medskip

Thus, Lemma \ref{N4colorimplyN5color} is shown.
\qed

\medskip

This finally leads to a polynomial number of feasible $xy$-colorings for each component in $G[N_2 \cup N_3 \cup N_4 \cup N_5]$ (recall that all these components can be independently colored). This leads to:

\begin{theorem}\label{S115frxydim}
If $|N_2| \geq 5$ and $N_5 \neq \emptyset$ then one can check in polynomial time whether $G$ has a d.i.m.\ containing $xy$.
\end{theorem}


This completes the proof that DIM can be solved for $S_{1,1,5}$-free graphs in polynomial time.

\medskip

\noindent
{\bf Acknowledgment.}
We are grateful to the anonymous referees for their helpful comments.
The second author would like to witness that he just tries to pray a lot and is not able to do anything without that - ad laudem Domini.

\begin{footnotesize}

\end{footnotesize}

\end{document}